\documentclass[12pt,onecolumn,draft]{IEEEtran}
\usepackage{amsmath,amssymb,amsfonts,latexsym,cite}
\usepackage[final]{graphicx}
\usepackage[usenames]{color}

\def\figref#1{Fig.~\ref{#1}}
\def\secref#1{Section~\ref{#1}}

\def\thmref#1{Theorem~\ref{#1}}
\def\corref#1{Corollary~\ref{#1}}
\def\expref#1{Example~\ref{#1}}

\def\bdA{\boldsymbol{A}}
\def\bdB{\boldsymbol{B}}

\def\bdH{\boldsymbol{H}}

\def\bdT{\boldsymbol{T}}

\def\bdV{\boldsymbol{V}}

\def\bda{\boldsymbol{a}}

\def\bdd{\boldsymbol{d}}
\def\bde{\boldsymbol{e}}

\def\bdh{\boldsymbol{h}}

\def\bdw{\boldsymbol{w}}
\def\bdx{\boldsymbol{x}}
\def\bdy{\boldsymbol{y}}
\def\bdz{\boldsymbol{z}}

\def\mC{\mathcal{C}}
\def\mD{\mathcal{D}}

\def\mM{\mathcal{M}}

\def\mP{\mathcal{P}}

\def\mS{\mathcal{S}}

\def\mV{\mathcal{V}}

\def\bdzeros{\boldsymbol{0}}
\def\bdones{\boldsymbol{1}}

\def\bdLambda{\boldsymbol{\Lambda}}

\def\CN#1{\mathcal{CN}(#1)} 

\def\diag{\text {diag}}

\def\E{\text{E}}

\def\tot{\textsf{total}}

\def\td{\tilde}

\def\bydef{:=}

\def\R#1{R_{#1}}

\long\def\comment#1{}
\newtheorem{remark}{Remark}
\newtheorem{theorem}{Theorem}

\newtheorem{thm}{Theorem}
\newtheorem{cor}[thm]{Corollary}

\newtheorem{example}{Example}

\makeatletter
\def\subsubsection{
     \@startsection {subsubsection}{3}{\z@ }%
     {-1.25ex\@plus -1ex \@minus -.2ex}%
     {0.6ex \@plus .4ex \@minus .1ex}%
     {\itshape}%
} \makeatother

\begin{document}

\title{Degrees of Freedom Region for an Interference  Network with General
 Message Demands}

\author{
Lei~Ke,  Aditya~Ramamoorthy,
\IEEEmembership{Member,~IEEE},\\ Zhengdao~Wang, \IEEEmembership{Senior
Member,~IEEE}, and Huarui~Yin, \IEEEmembership{Member,~IEEE}
\thanks{L. Ke was with the Department of
Electrical and Computer Engineering, Iowa State University, Ames, IA 50011 USA. He is now with Qualcomm Inc., San Diego, CA 92121
(e-mail: lke@qualcomm.com).}
\thanks{A. Ramamoorthy (corresponding author) and Z. Wang are with the Department of
Electrical and Computer Engineering, Iowa State University, Ames, IA 50011 USA
(e-mail: \{adityar, zhengdao\}@iastate.edu).}
\thanks{H. Yin is with the Department of Electronics Engineering and
Information Science, University of Science and Technology of China, Hefei,
Anhui, 230027, P.R.China (e-mail: yhr@ustc.edu.cn). He was visiting Iowa State
University when this work was performed.} \thanks{The material in this work was presented in
part at the 2011 IEEE International Symposium on Information Theory. This work was supported in part by NSF grant CCF-1018148.}}

\maketitle

\begin{abstract}
We consider a single hop interference network with $K$ transmitters
and $J$ receivers, all having $M$ antennas. Each transmitter emits
an independent message and each receiver requests an arbitrary
subset of the messages. This generalizes the well-known $K$-user
$M$-antenna interference channel, where each message is requested by
a unique receiver. For our setup, we derive the degrees of freedom
(DoF) region. The achievability scheme generalizes the interference
alignment schemes proposed by Cadambe and Jafar. In particular, we
achieve general points in the DoF region by using multiple base
vectors and aligning all interferers at a given receiver to the
interferer with the largest DoF. As a byproduct, we obtain the DoF
region for the original interference channel. We also discuss
extensions of our approach where the same region can be achieved by
considering a reduced set of interference alignment constraints,
thus reducing the time-expansion duration needed. The DoF region for
the considered system depends only on a subset of receivers whose
demands meet certain characteristics. The geometric shape of the DoF
region is also discussed.

\begin{IEEEkeywords}
Interference alignment, degrees of freedom region, multicast,
multiple-input multiple-output, interference network.
\end{IEEEkeywords}
\end{abstract}

\newpage
\linespread{1.6}\normalsize \newcommand{\figwidth}{\linewidth}

\section{Introduction}

In wireless networks, receivers need to combat interference from undesired
transmitters in addition to the ambient noise. Interference alignment has
emerged as an important technique in the study of fundamental limits of such
networks \cite{mamk06c,jash08}. Traditional efforts in dealing with
interference have focused on reducing the interference power, whereas in
interference alignment the focus is on reducing the dimensionality of the
interference subspace. The subspaces of interference from several undesired
transmitters are \emph{aligned} so as to minimize the dimensionality of the
total interference space. For the $K$-user $M$-antenna interference channel,
it is shown that alignment of interference is simultaneously possible at all the receivers, allowing each user to transmit at approximately half the
single-user rate in the high signal-to-noise ratio (SNR) scenario
\cite{caja08}. The idea of interference alignment has been successfully
applied to other interference networks as well \cite{caja09, suts08c, wesk07c,
gojw09, mali09a}.

The vector interference alignment schemes of \cite{caja08} are applicable to
time-varying channels. Constant channels have been dealt with using the
technique of real interference alignment
\cite{gojw09,brpt10,sjvj08c,etor09,mgmk09}. The major difference between
vector interference alignment and real interference alignment is that the
former relies on the linear vector-space independence, while the latter relies
on linear rational independence. Besides vector and real interference
alignment schemes, it is also possible to utilize the ergodicity of the
channel states in the so called ergodic interference alignment scheme
\cite{njgv09c}.

A majority of systems considered so far for interference alignment involve
only multiple unicast traffic, where each transmitted message is only demanded
by a single receiver. However, there are wireless \emph{multicast}
applications where a common message may be demanded by multiple receivers,
e.g., in a wireless video broadcasting. Such general message request sets have
been considered in \cite{ngjv09c} where each message is assumed to be
requested by an equal number of receivers. Ergodic interference alignment was
employed to derive an achievable sum rate. A different but related effort is
the study of the compound multiple-input single-output broadcast channel
\cite{gojw09,mali09a}, where the channel between the base station and the
mobile user is drawn from a known discrete set. As pointed out in
\cite{gojw09}, the compound broadcast channel can be viewed as a broadcast
channel with common messages, where each message is requested by a group of
receivers. Therefore, its total degrees of freedom (DoF) is also the total DoF
of a broadcast channel with different multicast groups. It is shown that using
real interference alignment scheme, the outer bound of the compound broadcast
channel \cite{wesk07c} can be achieved regardless of the number of channel
states one user can have. The compound setting was also explored for the $X$
channel and the interference channel in \cite{gojw09}, where the total number
of DoF is shown to be unchanged for these two channels. However, the DoF
region was not identified in \cite{gojw09}.

In this paper, we consider a natural generalization of the multiple unicasts
scenario considered in the work of Cadambe and Jafar \cite{caja08}. We
consider a setup where there are $K$ transmitters and $J$ (that may be
different from $K$) receivers, each having $M$ antennas. Each transmitter
emits a unique message and each receiver is interested in an arbitrary subset
of the $K$ messages. That is, we consider interference networks with general
message demands. Our main result in this paper is the DoF region for such
networks. One main observation is that by appropriately modifying the
achievability schemes of \cite{caja08,caja09}, we can achieve any point in the
DoF region. To the best of our knowledge, the DoF region in this scenario has not
been obtained before. Our main contributions can be summarized as follows
\begin{itemize}
\item[(i)] We completely characterize the DoF region for interference networks
with general message demands. We achieve any point in the DoF region by using
multiple base vectors and aligning the interference at each receiver to its
largest interferer. The geometric shape of the region is also discussed.

\item [(ii)] As a corollary, we obtain the DoF region for the case of multiple
unicasts considered in \cite{caja08}. We also provide an additional proof
based on timesharing for this case.

\item[(iii)] We discuss extensions of our approach where the DoF region can be
achieved by considering fewer interference alignment constraints, allowing for
interference alignment over a shorter time duration. We show that the region
depends only on a subset of receivers whose demands meet certain
characteristics.
\end{itemize}

This paper is organized as follows. The system model is given in
\secref{sec.model}. We present the DoF region of this system, and establish
its achievability and converse in \secref{sec.general.multicast}. We discuss
the approaches for reducing the number of alignment constraints, the DoF
region for the $K$-user $M$-antenna interference channel of \cite{caja08}, the
total DoF in \secref{sec.discuss}. Finally, \secref{sec.con} concludes our
paper.

We use the following notation: boldface uppercase (lowercase)
letters denote matrices (vectors). Real, integer, and complex
numbers sets are denoted by $\mathbb{R}$, $\mathbb{Z}$ and
$\mathbb{C}$, respectively. We define $\mathbb R^K_+ \bydef \{ (x_1,
x_2, \ldots, x_K): x_k\in \mathbb R, x_k \ge 0, 1\le k\le K \}$, and
define $\mathbb Z^K_+$ similarly. We use $\CN{0,1}$ to denote the
circularly symmetric complex Gaussian (CSCG) distribution with zero
mean and unit variance. For a vector $\bda$, $[\bda]_p$ is the $p$th
entry. For two matrices $\bdA$ and $\bdB$, $\bdA \prec \bdB$ implies
that the column space of $\bdA$ is a subspace of the column space of
$\bdB$.

\section{ System Model} \label{sec.model}

Consider a single hop interference network with $K$ transmitters and $J$
receivers. Each transmitter has one and only one independent message. For this
reason, we do not distinguish between the indices for messages and that for
transmitters. Each receiver can request an arbitrary set of messages from
multiple transmitters. Let $\mM_j$ be the set of indices of those messages
requested by receiver $j$. We assume that all the transmitters and receivers
have $M$ antennas. The channel between transmitter $k$ and receiver $j$ at
time instant $t$ is denoted as $\bdH_{jk}(t)\in \mathbb{C}^{M\times M}, 1\leq
k \leq K, 1\leq j \leq J$. We assume that the elements of all the channel
matrices at different time instants are independently drawn from some
continuous distribution. In addition, the channel gains are bounded between a
positive minimum value and a finite maximum value to avoid degenerate channel
conditions. The received signal at the $j$th receiver can be expressed as
\begin{align*}
\bdy_j(t)=\sum_{k=1}^K \bdH_{jk}(t)\bdx_k(t)+\bdz_j(t),
\end{align*}
where $\bdz_j\in \mathbb{C}^{M}$ is an independent CSCG noise with each entry
$\CN{0,1}$ distributed, and $\bdx_k(t)\in \mathbb{C}^{M}$ is the transmitted
signal of the $k$th transmitter satisfying the following power constraint
\begin{align*}
\E(||\bdx_k(t)||^2)\leq P, \quad 1\leq k \leq K.
\end{align*}
Henceforth, we shall refer to the above setup as an interference network with
general message demands. Our objective is to study the DoF region of an
interference network with general message demands when there is perfect CSI at
receivers and global CSI at transmitters. Denote the capacity region of such a
system as $C(P)$. The corresponding DoF region is defined as
\begin{align*}
\mD \bydef \left\{\bdd=(d_1,d_2,\cdots,d_{K})\in \mathbb R^K_+: \exists
(\R1(P),\R2(P),\cdots,R_K(P))\in C(P),  \right.\\
\left.  \text{such that } d_k
=\lim_{P\to\infty}\frac{R_k(P)}{\log(P)}, \quad 1\leq k \leq K \right \}.
\end{align*}

If $J=K$ and $\mM_j=\{j\}, \forall j$, the general model we considered here
will reduce to the well-known $K$ user $M$ antenna interference channel as in
\cite{caja08}.

\section{DoF Region of Interference Network with General Message Demands}
\label{sec.general.multicast}

In this section, we derive the DoF region of the interference network with
general message demands. Our main result can be summarized as the following
theorem.

\begin{theorem}\label{thm.main}
The DoF region of an interference network with general message demands with
$K$ transmitters, $J$ receivers, and $M$ antennas is given by
\begin{align}\label{eq.main}
\mD=\left\{\bdd\in\mathbb R^K_+: \sum_{k\in
\mM_j}d_k+\max_{i\in\mM_j^c}(d_i)\leq M,\quad \forall 1\leq j \leq J
 \right\},
\end{align}
where $\mM_j$ is the set of indices of messages requested by receiver $j$,
$1\leq j \leq J$. \hfill\QED
\end{theorem}

\subsection{Discussion on the DoF region}

\subsubsection{The converse argument}

To show the region given by \eqref{eq.main} is an outer bound, we use a genie
argument which has been used in several previous papers, e.g., \cite{jafa07,
caja08}. In short, we assume that there is a genie who provides all the
interference messages except for the interference message with the largest DoF
to receiver $j$. Thus, receiver $j$ can decode its intended messages,
following which it can subtract the intended message component from the
received signal so that the remaining interfering message can also be decoded.
Hence, \eqref{eq.main} follows due to the multiple access channel outer bound.

\begin{figure}
\centerline{\includegraphics[width=0.3 \figwidth]{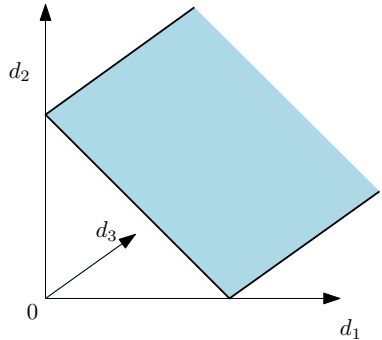}}
\caption{The cylinder set defined by $d_1+d_2\le M$ and $d_i\ge 0$, $i=1,2$,
in a three-dimensional space.}
\label{fig.cylinder}
\end{figure}

\subsubsection{Geometric Shape of the DoF region} The DoF region is a convex
polytope, as is evident from the representation in \eqref{eq.main}. The
inequalities in \eqref{eq.main} characterize the polytope as the intersection
of half spaces, each defined by one inequality. Note that all the coefficients
of the DoF terms in each inequality are either zero or one. That is, all the
inequalities are of the form
\begin{equation}\label{eq.simplex}
  \sum_{i\in {\mS}} d_i \le M
\end{equation}
where $\mS$ is a subset of $\{1, 2, \ldots, M\}$. This can be seen by
expanding each inequality in \eqref{eq.main} containing a ``max'' term into
several inequalities that do not contain the maximum operator. For example, we
can expand $d_1+\max(d_2, d_3)\le M$ into $d_1+d_2\le M$ and $d_1+d_3\le M$.
In a $|\cal S|$-dimensional space, the set of points defined by $\sum_{i\in
{\mS}} d_i = M$ and $d_i\ge 0, i\in \mS$ is a simplex of $(|\mS|-1)$
dimensions. For example, $d_1+d_2=M$ describe a one-dimensional simplex. This
simplex, together with the lines (planes) $d_1=0$ and $d_2=0$ defines a subset
of the 2-dimensional space, which is a right triangle of equal sides. When
considering such an inequality in the $K$-dimensional space, each such
inequality describes a cylinder set whose projection into the
$|\mS|$-dimensions is the aforementioned subset enclosed by the simplex and
the planes $d_i=0, i\in \mS$. See \figref{fig.cylinder} for an illustration in
the case of $K=3$ and $\mS=\{1,2\}$. The whole DoF region therefore is the
intersection of such cylinder sets.

It is also possible to specify convex polytopes via its vertices.
Theoretically it is possible to find all the vertices of the DoF region by
solving a set of linearly independent equations, by replacing a subset of $K$
inequalities to equalities, and verifying that the solution satisfies all
other constraints. However the number of such equations can be as large as
$\binom{J(K-1)+K}{K}$, where $J(K-1)+K$ is the total number of (expanded)
inequalities. Nevertheless, in some special cases as we will see later, it is
possible to find the vertices exactly.

In the following part, we will use a simple example to demonstrate the DoF
region and reveal the basic idea of our achievability scheme.

\begin{figure}
\centerline{\includegraphics[width=0.7
\figwidth]{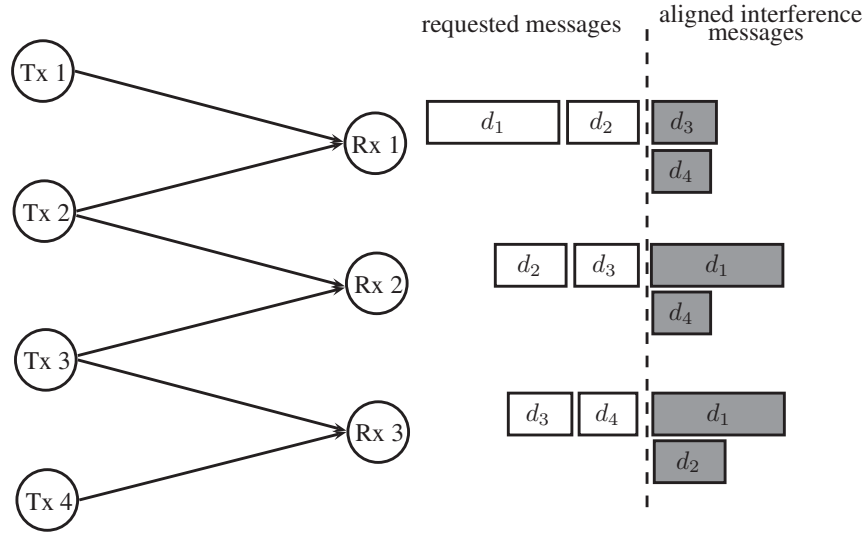}} \caption{Left: Example system
with arrows denoting the demands. Right: Alignment scheme for
achieving DoF point $(d_1,d_2,d_3, d_4)=(1-2d_{4},d_{4}, d_{4},
d_{4})$.}

\label{fig.dofExample_channel}
\end{figure}

\subsection{An example of the general message demand and the DoF region}

We first show the geometric picture of the DoF region for a specific example,
which is useful for developing the general achievability scheme.

\begin{figure}
\centerline{\includegraphics[width=0.9 \figwidth]{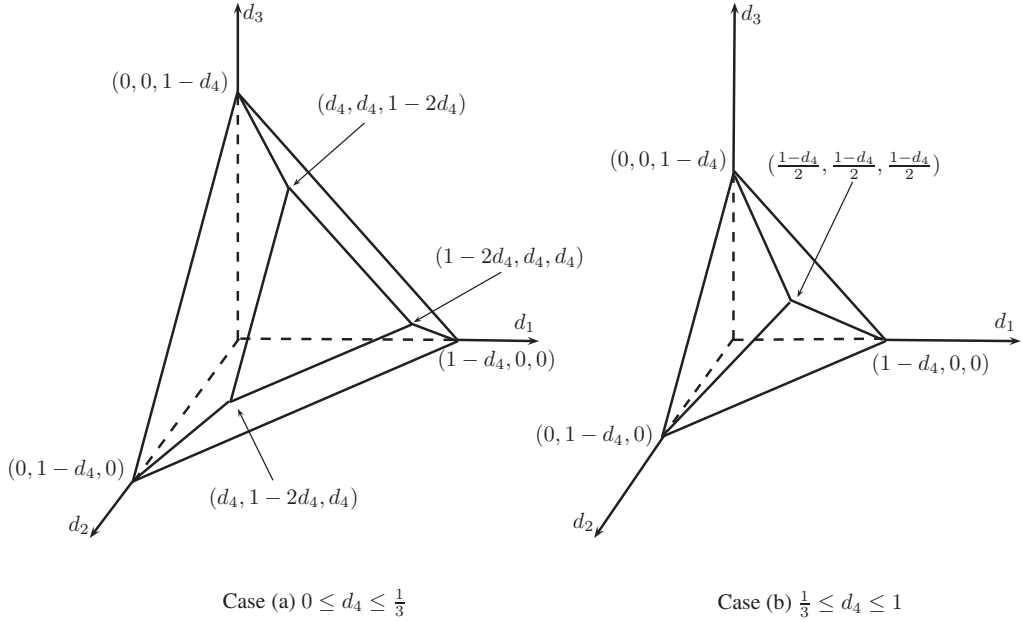}}
\caption{DoF region in lower dimensions as a function of $d_4$.}
\label{fig.dofRegionExample}
\end{figure}

Consider an interference network with $4$ transmitters and $3$ receivers; see
\figref{fig.dofRegionExample}. All the transmitters and receivers have single
antenna; that is, $M=1$. Assume $\mM_1=\{1,2\}$, $\mM_2=\{2,3\}$ and
$\mM_3=\{3,4\}$. The DoF region of the system according to
Theorem~\ref{thm.main} is as follows
\begin{align}
\mD=\left\{\bdd\in\mathbb R^4_+\left|\begin{array}{c}
d_1+d_2+d_3\leq1 \\
d_1+d_2+d_4\leq1 \\
d_2+d_3+d_4\leq1 \\
d_1+d_3+d_4\leq1 \\
\end{array} \right.
 \right\}. \label{eq.dofregionexample}
\end{align}

The region is 4-dimensional and hence difficult to illustrate.
However, if the DoF of one message, say $d_4$, is fixed, the DoF
region of the other messages can be illustrated in lower dimensions
as a function of $d_4$; see \figref{fig.dofRegionExample}.

We first investigate the region when $0\leq d_4 \leq \frac{1}{3}$, for which
the coordinates of the vertices are given in \figref{fig.dofRegionExample},
case (a). The achievability of the vertices on the axes is simple as there is
no need of interference alignment. Time sharing between the single-user rate
vectors $\{\bde_k, k=1, 2, \ldots, K\}$ is sufficient. For the remaining three
vertices, we only need to show the achievability of one point as the
achievability of the others are essentially the same by swapping the message
indices.

We will use the scheme based on \cite{caja08} to do interference alignment and
show $(d_1,d_2,d_3, d_4)=(1-2d_{4},d_{4}, d_{4}, d_4)$ is achievable for any
$0\le d_4\le \frac 13$. Let $\tau$ denote the
duration of the time expansion in number of symbols. Here and after, we use
the superscript tilde $\tilde{(\cdot)}$ to denote the time expanded signals, e.g.,
 $\tilde \bdH_{jk}=\diag(\bdH_{jk}(1), \bdH_{jk}(2), \dots,
\bdH_{jk}(\tau))$, which is a size $\tau\times \tau$ diagonal matrix (recall
that $M=1$). Denote the beamforming matrix of transmitter $k$ as $\tilde
\bdV_k$. First, we want messages 3 and 4 to be aligned at receivers 1. Notice
that messages 3 and 4 have the same number of DoF. We choose to design
beamforming matrices such that the interference from transmitter 4 is aligned
to interference from transmitter 3 at receiver 1. Therefore we have the
following constraint
\begin{align}
\td \bdH_{14}\td \bdV_4&\prec\td \bdH_{13}\td \bdV_3. \label{eq.regionnexp.rx1}
\end{align}
Note that the interference due to transmitter 1 has a larger DoF at receiver
2; thus, we must align interference from transmitter 4 to interference from
transmitter 1 at receiver 2, which leads to
\begin{align}
\td \bdH_{24}\td \bdV_4&\prec\td \bdH_{21}\td \bdV_1. \label{eq.regionnexp.rx2}
\end{align}
Similarly at receiver 3, we have
\begin{align}
\td \bdH_{42}\td \bdV_2&\prec\td \bdH_{41}\td \bdV_1. \label{eq.regionnexp.rx3}
\end{align}

The alignment relationship is also shown in \figref{fig.dofExample_channel}.
Notice that $d_1$ is larger than $d_2$, $d_3$ and $d_4$. Therefore it is
possible to design $\td\bdV_1$ into two parts as
$[\td\bdV_{1a},\td\bdV_{1b}]$, where $\td\bdV_{1a}$ is used for transmitting
part of the message 1 with the same DoF as other messages. The second part
$\td\bdV_{1b}$ is used for transmitting the remaining DoF of message 1. In
addition, all the columns in $[\td\bdV_{1a},\td\bdV_{1b}]$ are linearly
independent.

The design of $\td\bdV_{1a}$ can be addressed by the classic asymptotic
interference alignment scheme in \cite{caja08}. The beamforming matrices in
\cite{caja08} are chosen from a set of beamforming columns, whose elements are
generated from the product of the powers of certain matrices and a vector. We
term such a vector as a \emph{base vector} in this paper. The base vector was
chosen to be the all-one vector in \cite{caja08}. The scheme proposed in
\cite{caja08} was further explored for wireless $X$ network \cite{caja09} with
multiple independent messages at single transmitter, where multiple
independent and randomly generated base vectors are used for constructing the
beamforming matrices. In our particular example, as no interference is aligned
to the second part of message 1, we may choose an independent and randomly
generated matrix for $\td\bdV_{1b}$. However, in general we need to
construct the beamforming matrices in a structured manner using multiple base vectors as
we will see in \secref{sec.ach.same.dof.single}. The DoF point can be achieved
asymptotically when the number of time expansion $\tau$ goes to infinity. We
omit further details of beamforming construction for this particular example.

The DoF region of case (b) in \figref{fig.dofRegionExample} can be
achieved similarly by showing that the vertex $(d_1,d_2,d_3,
d_4)=(\frac{1-d_4}{2},\frac{1-d_4}{2},\frac{1-d_4}{2}, d_4)$ is achievable.
This also requires the multiple base vector technique.

We remark that the DoF region in this example can also be formulated as the
convex hull of the following vertices $\{\bdzeros, \bde_1, \bde_2, \bde_3,
\bde_4, \frac 13 \bdones\}$. The achievability of the whole DoF therefore can
be alternatively established by showing that $\frac 13 \bdones=(\frac 13,
\frac 13, \frac 13, \frac 13)$ is achievable. This can be verified by exhaustively examining the basic feasible solutions for the polytope description in \eqref{eq.dofregionexample}.

\subsection{Achievability of the DoF region with single antenna transmitters
and receivers} \label{sec.ach.same.dof.single} We first consider the
achievability scheme when all the transmitters and receivers have a single
antenna, i.e., $M = 1$.  It is evident that we only need to show any point
in $\mD$ satisfying
\begin{align}
d_K\leq d_{K-1}\leq\dots \leq d_2\leq d_1 \label{eq.relation.di}
\end{align}
is achievable, for otherwise the messages can be simply renumbered so that
$\eqref{eq.relation.di}$ is true.

\subsubsection{The set of alignment constraints} The achievability scheme is
based on interference alignment over a time expanded channel. Based
on \eqref{eq.relation.di}, we impose the following relationship on
the sizes of the beamforming matrices of the transmitters:
\begin{align}
|\tilde\bdV_K|\leq |\tilde\bdV_{K-1}|\leq \dots\leq
  |\tilde\bdV_2|\leq |\tilde\bdV_1|,
\label{eq.relation.vi}
\end{align}
where $|\bdV|$ denotes the number of columns of matrix $\bdV$. At receiver $j$,
we always align the interference messages with larger indices to the
interference message with index $\delta_j$, which is the interference message
with the largest DoF, given as
\begin{align*}
\delta_j=\min\{k|k\in \mM_j^c\}.
\end{align*}
Denote $\bdT_{m,n}^{[j]}$ as following
\begin{align*}
\bdT_{m,n}^{[j]}=\td\bdH_{jm}^{-1}\td\bdH_{jn},
\end{align*}
which is the matrix corresponding to the alignment constraint
\begin{align*} \td\bdH_{jn}\td\bdV_n \prec \td\bdH_{jm}
\td\bdV_m,
\end{align*}
that enforces the interference from message $n$ to be aligned to the
interference of message $m$ at receiver $j$. Based on \eqref{eq.relation.vi},
for any $\bdT_{m,n}^{[j]}$ matrix, we always have $n>m$.

For convenience, we define the following set
\begin{align}
\label{eq.constraintset}
\mC&\bydef \left\{(m,n,j)\left|\begin{array}{c}
 j \in \{1,\dots,J\},\\ m,n\in \mM_j^c, \\
m=\delta_j, n>m
\end{array} \right.  \right\}.
\end{align}
In other words, $\mC$ is a set of vectors denoting all the alignment
constraints. There exists a one-to-one mapping from a vector $(m,n,j)$ in
$\mC$ to the corresponding matrix $\bdT_{m,n}^{[j]}$.

\subsubsection{Time expansion and base vectors} It is not difficult to see
that the vertices of the DoF region given in \eqref{eq.main} must be rational
as all the coefficients and right hand side bounds are integers (either zero
or one). Therefore we only need to consider the achievability of such rational
vertices, although the proof below applies to any interior rational points in
the DoF region as well.

For any rational DoF point $\bdd$ within $\mD$ (vertex or not) satisfying
\eqref{eq.relation.di}, we can choose a positive integer $\kappa$, such that
\begin{align}
\kappa\bdd&=(\bar d_1,\bar d_2,\dots,\bar d_K)\in\mathbb{Z}_+^K.
\label{eq.kappa.integer}
\end{align}

\begin{figure} \centerline{\includegraphics[width=0.9
\figwidth]{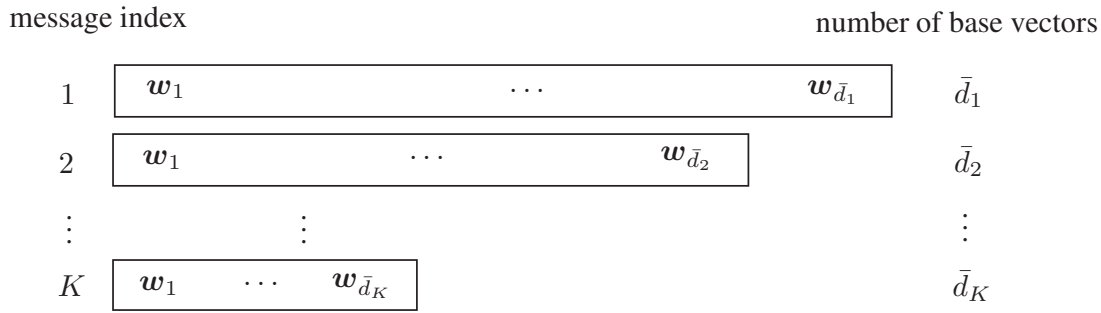}}
\caption{Illustration of the base vectors used by different messages. The base vectors used by transmitter $k$ will also be used by transmitters $1, \dots, k-1$.}
\label{fig.basevec}
\end{figure}

We then use multiple base vectors to construct the beamforming matrices. The
total number of base vectors is $\bar d_1$. Denote the base vectors as
$\{\bdw_i, 1\leq i \leq \bar d_1\}$. Transmitter $k$ will use base vectors
$\bdw_i, 1\leq i \leq \bar d_k$ to construct its beamforming matrix, and the
same base vectors will be used by transmitters $1, 2, \ldots, k-1$ as well
(see \figref{fig.basevec}). The elements of $\bdw_i$ are independent and
identically drawn from some continuous distribution. In addition, we assume
that the absolute value of the elements of $\bdw_i$ are bounded between a
positive minimum value and a finite maximum value, in the same way that
entries of $\bdH_{jk}(t)$ are bounded (see \secref{sec.model}).
  Denote $\Gamma=|\mC|$, which
is the total number of $\bdT_{m,n}^{[j]}$ matrices as well. We propose to use
a $\tau=\kappa(l+1)^\Gamma$ fold time expansion, where $l$ is a positive
integer.

\subsubsection{Beamforming matrices design}

The beamforming matrices are generated in the following manner.
\begin{enumerate}
 \item[i)] Denote $\Gamma_k$
as the cardinality of the following set
\begin{align*}
\mC_k=\left\{(m,n,j)| (m,n,j) \in \mC , n\leq k \right\} \quad k=1,2\dots,K,
\end{align*}
which is the number of matrices whose exponents are within
$\{0,1,\dots,l-1\}$, while the other $\Gamma-\Gamma_k$ matrices can be raised
to the power of $l$. It is evident that $\Gamma_K=\Gamma$, and $\Gamma_1=0$.

\item[ii)] Transmitter $K$ uses $\bar d_K$ base vectors. For base vector
$\bdw_i, 1\leq i\leq \bar d_K$, it generates the following $l^\Gamma$ columns
\begin{align*}
\prod_{(m,n,j)\in \mC}\left(\bdT_{m,n}^{[j]}\right)^{\alpha_{m,n,j}}\bdw_i
\end{align*}
where $\alpha_{m,n,j}\in\{0,1,\dots,l-1\}$. Hence, the total number of columns
of $\td \bdV_K$ is
 $\bar d_Kl^\Gamma$.

\item[iii)] Similarly, transmitter $k$ uses $\bar d_{k}$ base vectors. For
base vector $\bdw_i, 1\leq i\leq \bar d_k$, it generates
$l^{\Gamma_{k}}(l+1)^{\Gamma-\Gamma_{k}}$ columns\begin{align}
\prod_{(m,n,j)\in \mC}\left(\bdT_{m,n}^{[j]}\right)^{\alpha_{m,n,j}}\bdw_i
\label{eq.bfcolumnkgel}
\end{align}
where
\begin{align*}
\alpha_{m,n,j}&\in\{0,1,\dots,l\}& n>k\\
\alpha_{m,n,j}&\in\{0,1,\dots,l-1\}& n\leq k
\end{align*}
 \end{enumerate}

In summary, the beamforming design is as follows, for every message, we
construct a beamforming column set as \begin{align*} \td \mV_k&\!=\!\left\{\!
  \prod_{(m,n,j)\in \mC}\!
    \left(\bdT_{m,n}^{[j]}\right)^{\!\alpha_{m,n,j}}\!\!\bdw_i\bigg|
 1\!\leq \!i\! \leq \!\bar d_k,
 \alpha_{m,n,j}\!\in\!\begin{cases}\{0,1,\dots, l\} & \text{\!\!if
 } n> k
 \\
\{0,1,\dots, l-1\} & \text{\!\!otherwise} \\
\end{cases} \right\}\quad \!1\!\leq\! k\!\leq\! K.
\end{align*}
The beamforming matrix $\td\bdV_k$ is chosen to be the matrix that contains
all the columns of $\td \mV_k$.

\subsubsection{Alignment at the receivers}

Assume $(k, k', j)\in \mC$, so that message $k'$ needs to be aligned with
message $k<k'$ at receiver $j$. We now show that this is guaranteed by our
design. Let $\bdw_i$, $1\le i\le \bar d_{k'}$ be a base vector used by
transmitter $k'$, and hence also used by transmitter $k$. From
\eqref{eq.bfcolumnkgel}, the beamforming vectors generated by $\bdw_i$ at
transmitter $k$ can be expressed in the following way
\begin{align}
\underbrace{
  \prod_{\substack{(m,n,j)\in \mC\\n \leq k}}
  \left(\bdT_{m,n}^{[j]}\right)^{\alpha_{m,n,j}}
  }_{\alpha_{m,n,j}\in\{0,1,\dots,l-1\}}
\underbrace{\prod_{\substack{(m,n,j)\in \mC\\
(m,n)=(k,k')}}\left(\bdT_{m,n}^{[j]}\right)^{\alpha_{m,n,j}}
  \prod_{\substack{(m,n,j)\in \mC\\ n>k\\ (m,n)\ne (k,k')}}
  \left(\bdT_{m,n}^{[j]}\right)^{\alpha_{m,n,j}}}
  _{\alpha_{m,n,j}\in\{0,1,\dots,l\}}\bdw_i
\label{eq.bfcolumnk},
\end{align}
whereas those at the transmitter $k'$ can be expressed as
\begin{align}
\underbrace{
  \prod_{\substack{(m,n,j)\in \mC\\n \le k'\\(m,n)\ne(k,k')}}
  \left(\bdT_{m,n}^{[j]}\right)^{\alpha_{m,n,j}}
  \prod_{\substack{(m,n,j)\in \mC\\(m,n)=(k,k')}}
  \left(\bdT_{m,n}^{[j]}\right)^{\alpha_{m,n,j}}}
  _{\alpha_{m,n,j}\in\{0,1,\dots,l-1\}}
\underbrace{\prod_{\substack{(m,n,j)\in \mC\\n>k'}}
  \left(\bdT_{m,n}^{[j]}\right)^{\alpha_{m,n,j}}}
  _{\alpha_{m,n,j}\in\{0,1,\dots,l\}}\bdw_i.
\label{eq.bfcolumnkp1}
\end{align}

Comparing the ranges of $\alpha_{k,k',j}$ in \eqref{eq.bfcolumnk} and
\eqref{eq.bfcolumnkp1}, i.e., the middle terms, it can be verified that the
columns in \eqref{eq.bfcolumnkp1} multiplied with $\bdT_{k,k'}^{[j]}$, will be
a column in \eqref{eq.bfcolumnk}, $\forall (k,k',j)\in \mC$. That is, message
$k'$ can be aligned to message $k$ for any $j$ such that $(k,k',j)\in \mC$.

The alignment scheme works due to the following reasons.
\begin{enumerate}
\item[i)] Let $\alpha_{m,n,j}(k)$ denote the exponent of the $(m,n,j)$ term
for $\td \bdV_k$. The construction of the beamforming column set guarantees
that
\begin{align}
 \max \alpha_{m,n,j}(m) > \max \alpha_{m,n,j}(n),
  \quad \forall(m,n,j) \in \mC
\label{eq.max}
\end{align}
 by setting
\begin{align*}
 \max \alpha_{m,n,j}(m)&=l,\\
 \max \alpha_{m,n,j}(n)&=l-1.
\end{align*}
With \eqref{eq.max}, we are guaranteed all vectors in $\td\bdV_n$, when left
multiplied with $\bdT_{m,n}^{[j]}$ (which has the effect of increasing the
exponent of $\bdT_{m,n}^{[j]}$ by one), generates a vector that is within the
columns of $\td\bdV_m$. Hence the alignment is ensured.

For other terms where $k$ is not $m$ or $n$, $\max \alpha_{m,n,j}(k)$ can be
either $l$ or $l-1$.
\item[ii)] The base vectors used by transmitter $n$ are also used by
transmitter $m<n$. This guarantees that if the interference from transmitter
$n$ needs to be aligned with interference from transmitter $m$, where $m<n$,
the alignment is ensured with the condition \eqref{eq.max}.
\end{enumerate}

\subsubsection{Achievable Rates}

It is evident that $\td\bdV_k$ is a tall matrix of dimension
${\kappa(l+1)^\Gamma}\times \bar d_kl^{\Gamma_k}(l+1)^{\Gamma-\Gamma_k}$. We
also need to verify it has full column rank. Notice that all the entries in
the upper square sub-matrix are monomials and the random variables of the
monomial are different in different rows. In addition, for a given row $r,
1\leq r \leq \bar d_kl^{\Gamma_k}(l+1)^{\Gamma-\Gamma_k}$, any two entries
have different exponents. Therefore, based on \cite[Lemma 1]{caja09},
$\td\bdV_k$ has full column rank and
\begin{align*}
\lim_{l\rightarrow \infty}\frac{|\td\bdV_k|}{\tau}=\lim_{l\rightarrow \infty}
\frac{\bar d_kl^{\Gamma_k}(l+1)^{\Gamma-\Gamma_k}}{\kappa(l+1)^\Gamma}
=\frac{\bar d_{k}}{\kappa}=d_k.
\end{align*}

\comment {For any $(m',n',j')\in \mC$, the interference space of message $n'$
at receiver $j'$ can be expressed in the following form:
\begin{align}\label{eq.in}
\td \bdH_{j'n'}\left( \bdT_{m',n'}^{[j']} \right)^{\alpha_{m',n',j'}}\left
(\prod_{\substack{(m,n,j\in \mC)\\(m,n,j)\neq (m',n',j')}}
  \left(\bdT_{m,n}^{[j]}\right)^{\alpha_{m,n,j}} \right)\bdw_i:
  \quad\begin{cases} 1\leq i \leq \bar d_{n'} &  \\
0\leq\alpha_{m',n',j'}\leq l-1 &  \\
0\leq\alpha_{m,n,j}\leq l & n>n'\\
0\leq\alpha_{m,n,j}\leq l-1 & n<n' \\
\end{cases}.
\end{align}
The interference space of message $m'$ at receiver $j'$ can be expressed as
\begin{align}\label{eq.im}
\td \bdH_{j'm'}\left( \bdT_{m',n'}^{[j']} \right)^{\alpha_{m',n',j'}}\left
(\prod_{\substack{(m,n,j\in \mC)\\(m,n,j)\neq (m',n',j')}}
  \left(\bdT_{m,n}^{[j]}\right)^{\alpha_{m,n,j}} \right)\bdw_i:
  \quad\begin{cases} 1\leq i \leq \bar d_{m'} &  \\
0\leq\alpha_{m',n',j'}\leq l &  \\
0\leq\alpha_{m,n,j}\leq l & n>m'\\
0\leq\alpha_{m,n,j}\leq l-1 & n<m' \\
\end{cases}.
\end{align}
By the very definition of $\bdT_{m',n'}^{[j']}$, the vectors in \eqref{eq.in}
is a subset of the vectors in \eqref{eq.im}. Such alignment is true for any
$(m',n',j')$ in $\mC$.}

\subsubsection{Separation of the signal and interference spaces}

Finally, we need to ensure that the interference space and signal space are
linearly independent for all the receivers. Let the set of messages requested
by receiver $j$ be $\mM_j=\{m_{1,j},m_{2,j},\dots,m_{\beta_j,j}\}$, where
$\beta_j=|\mM_j|$. For receiver $j$ to be able to decode its desired messages,
the following matrix
\begin{align}
\bdLambda_j=\left[\td \bdH_{jm_{1,j}}\tilde \bdV_{m_{1,j}} | \td\bdH_{jm_{2,j}}
  \tilde \bdV_{m_{2,j}} | \dots,\td\bdH_{jm_{\beta_j,j}}
  \tilde \bdV_{m_{\beta_j,j}} | \td\bdH_{j\delta_j}\tilde \bdV_{\delta_j} \right]
  \label{eq.checkrank}
\end{align}
needs to have full rank for all $ 1\leq j \leq J$.

Notice that for any point within $\mD$
\begin{align}
\sum_{m\in\mM_j}\bar d_m+ \bar d_{\delta_j}\leq \kappa
\label{eq.dofbound.timexp}
\end{align}
always holds (recall $M=1$). Therefore $\bdLambda_j$ is a matrix
that is either tall or square. For any row $r$ of its upper square
sub-matrix, its elements can be expressed in the following general
form:
\begin{align*}
\bdH_{jk}(r)\prod_{(m,n,j)\in \mC}
  \left(\bdH_{jn}^{-1}(r)\bdH_{jm}(r) \right)^{\alpha_{m,n,j}}[\bdw_i]_r.
\end{align*}
The elements from different blocks (that is, different $[\td\bdH_{jk}\tilde
\bdV_k]$, $k\in\mM_j\cup\{\delta_j\}$) are different due to the fact that
$\bdH_{jk}(r)$'s are different, hence the monomials involve different sets of
random variables. Within one $\td\bdH_{jk}\tilde \bdV_k$, two monomials are
different either because they have different $[\bdw_i]_r, 1\leq i\leq \bar
d_k$, or, if they have the same $[\bdw_i]_r$, the associated exponents
$\alpha_{m,n,j}$ are different. Thus matrix $\bdLambda_j$ has the following
properties.
\begin{itemize}
\item[i)] Each term is a monomial of a set of random variables.
\item[ii)] The random variables associated with different rows are
independent.
\item[iii)] No two elements in the same row have the same exponents.
\end{itemize}
It follows from \cite[Lemma 1]{caja09} that $\bdLambda_j$ has full column rank
with probability one.

Combining the interference alignment and the full-rank arguments, we conclude
that any point $\bdd$ satisfying \eqref{eq.main} is achievable.

\subsection{Achievability of the DoF region with multiple antenna transmitters
and receivers} \label{sec.ach.same.dof.multi}

We next present the achievability scheme for the multiple antenna case. We
assume that all transmitters and receivers are equipped with the same number
$M$ of antennas. An achievability scheme optimal for the total DoF has been
proposed in \cite{caja08} based on an antenna splitting argument. However, the
same antenna splitting argument cannot be used to establish the DoF region in
general because it relies on the fact that the DoF's of the messages are
equal, which is the case when the total DoF is maximized. Indeed if one
attempts to perform antenna splitting with unequal DoF's and then applies the
previous scheme (Section \ref{sec.ach.same.dof.single}) by converting it into
a $MK \times MJ$ single antenna instance with independent messages at each
antenna, then the genie-based outer bound may rule out decoding at certain
receivers.

We now show the achievability of the DoF region of multiple antenna case based
on the method that was proposed in \cite{goja08a}. The messages are split at
the transmit side and transmitted via virtual single antenna transmitters,
while the receivers are still using all $M$ antennas to recover the intended
messages. Therefore, the one-to-many interference alignment scheme given in
\cite{goja08a} can be used here along with the multiple base vectors technique
to achieve the DoF region.

We assume that \eqref{eq.relation.di} is still true. After splitting the
transmitters, we now have an interference network with $MK$ virtual single antenna
transmitters and $J$ multiple antenna receivers. For any transmitter $k$, the
$p$th antenna will transmit a message of DoF $d_k/M$. In addition, the
beamforming matrices for all the virtual single antenna transmitters of
original system transmitter $k$ are the same, denoted as $\td\bdV_k$, and
therefore \eqref{eq.relation.vi} still holds. However, its size will be
different from the single antenna case as we will see in the discussion below.

\subsubsection{The set of alignment constraints} The channels in the modified
case are all in single input and multiple output representation. We denote the
channel between the $p$th antenna of transmitter $k$ and receiver $j$ as
$\bdh_{jk,p}$. Apparently,
$[\bdh_{jk,1},\bdh_{jk,2},\cdots,\bdh_{jk,M}]=\bdH_{jk}$. The channel
$\bdh_{jk,p}$ after time expansion is denoted as $\td\bdH_{jk,p}$, which is a
tall matrix
 of size $M\tau\times \tau$. At receiver $j$,
we still align the interference messages with larger indices to the
interference message with index $\delta_j$. However, because any $M$ channel
vectors from virtual single antenna transmitters to any receiver with $M$
antennas are linearly independent, it is impossible to align the interference
between only two virtual single antenna transmitters.  To achieve alignment at the receivers, we employ a design in
\cite{goja08a}, where the signal from one antenna is aligned with the signals
coming from \emph{all} the antennas of another transmitter. For our problem,
we will align at receiver $j$ the message from the $p$th antenna of
transmitter $n$ with the messages from all the antennas of transmitter
$\delta_j$, for all $n>\delta_j$ and for all $j$.
Specifically, letting $m=\delta_j$ for notational simplicity, we require
\begin{align}
\td\bdH_{jn,p}\td\bdV_n\prec \underbrace{[\td\bdH_{jm,1},\td\bdH_{jm,2},
  \cdots,\td\bdH_{jm,M}]}_{\td\bdH_{jm,1:M}}
\begin{bmatrix}\td\bdV_m & \bdzeros & \cdots & \bdzeros \\
\bdzeros & \td\bdV_m & \cdots & \bdzeros \\
\vdots & \bdzeros & \ddots & \vdots \\
\bdzeros & \bdzeros & \cdots & \td\bdV_m \\
\end{bmatrix} \label{eq.multi.antenna.align.diag}
\end{align}
The matrix $\td\bdH_{jm,1:M}$ is full rank and hence invertible. It is shown
in \cite{goja08a} that $\td\bdH_{jm,1:M}^{-1}\td\bdH_{jn,p}$ is an
$M\tau\times \tau$ matrix having block form
\begin{align*}
\td\bdH_{jm,1:M}^{-1}\td\bdH_{jn,p}=\begin{bmatrix}\bdT_{m,n,p,1}^{[j]} \\
\bdT_{m,n,p,2}^{[j]} \\
\vdots \\
\bdT_{m,n,p,M}^{[j]} \\
\end{bmatrix},
\end{align*}
where all block matrices $\bdT_{m,n,p,q}^{[j]}, 1\leq q\leq M$ are diagonal
 (see Appendix A in \cite{goja08a}) and therefore commutable. Hence, the
constraint \eqref{eq.multi.antenna.align.diag} can be converted to $M$
equivalent constraints:
\begin{align*}
\bdT_{m,n,p,q}^{[j]}\td\bdV_n\prec\td\bdV_m\qquad 1\leq q\leq M.
\end{align*}
Similar to the single antenna case, we define a set $\mC^M$ as follows
\begin{align*}
\mC^M&\bydef \left\{(m,n,p,q,j)\left|\begin{array}{c}
 j \in \{1,\dots,J\},\\ m,n\in \mM_j^c, m=\delta_j, n>m \\
1\leq p\leq M, 1\leq q\leq M
\end{array} \right.  \right\}.
\end{align*}
And there exists a one-to-one mapping from a vector $(m,n,p,q,j)$ in $\mC^M$
to the corresponding matrix $\bdT_{m,n,p,q}^{[j]}$. In addition, it is easy to
see that $|\mC^M|=M^2|\mC|$, where $\mC$ denotes the constraint set as defined
in \eqref{eq.constraintset} for the single antenna case.

\subsubsection{Time expansion and base vectors}

Similar to the single antenna case, we still need to use multiple base vectors
to construct the beamforming matrices. Recall $\kappa$ is a positive integer
such that \eqref{eq.kappa.integer} is still valid. The total number of base
vectors is still $\bar d_1$. For transmitter $k$, it uses base vector
$\bdw_{i}, 1\leq i \leq \bar d_k$ and all its antennas use all the base
vectors. Denote $\Gamma^M=|\mC^M|$. We propose to use $\tau=\kappa
M^2(l+1)^{\Gamma^M/M}$ fold time expansion.

\subsubsection{Beamforming matrices design}
The beamforming matrices can be generated in the following way

\begin{enumerate}
 \item[i)]
For any given $q$ where $1 \leq q \leq M$, denote $\Gamma_{q}^M$ as the
cardinality of the following set
\begin{align*}
\mC_{q}^M=\left\{(m,n,p,q,j)| (m,n,p,q,j) \in \mC^{M}, \forall m,n,p,j \right\}
\end{align*}
Furthermore, denote $\Gamma_{k,q}^M$ as the cardinality of the following set
\begin{align*}
\mC_{k,q}^M=\left\{(m,n,p,q,j)| (m,n,p,q,j) \in \mC^M , n\leq k \right\} \quad
1\leq k\leq K, 1 \leq q \leq M
\end{align*}
which is the number of matrices whose exponents are within
$\{(q-1)(l+1),(q-1)(l+1)+1,\dots,q(l+1)-2\}$, while the other
$\Gamma_{q}^M-\Gamma_{k,q}^M$ matrices can be raised to the power of up to
$q(l+1)-1$. It is evident that
\begin{align*}
\Gamma^M_q&=\Gamma^M/M,\quad \forall q\\
\Gamma^M_{K,q}&=\Gamma^M_q,\\
\Gamma^M_{1,q}&=0.
\end{align*}

\item[ii)] Transmitter $K$ uses $\bar d_K$ base vectors. For base vector $\bdw_i,
1\leq i\leq \bar d_K$, it generates the following $Ml^{\Gamma^M_q}$ columns
\begin{align*}
\bigcup_{1\leq q \leq M}\left\{ \prod_{(m,n,p,q,j)\in
\mC^M_q}\left(\bdT_{m,n,p,q}^{[j]}\right)^{\alpha_{m,n,p,q,j}}\bdw_i \right \}
\end{align*}
where $\alpha_{m,n,p,q,j}\in\{(q-1)(l+1),(q-1)(l+1)+1,\dots,q(l+1)-2\}$. Hence, the total number
of columns of $\td \bdV_K$ is
 $M\bar d_Kl^{\Gamma^M_q}$.

\item[iii)] Similarly, transmitter $k$ uses $\bar d_{k}$ base vectors. For base
vector $\bdw_i, 1\leq i\leq \bar d_k$, it generates
$Ml^{\Gamma_{k,q}^M}(l+1)^{\Gamma^M_q-\Gamma^M_{k,q}}$ columns\begin{align}
\bigcup_{1\leq q \leq M}\left\{\prod_{(m,n,p,q,j)\in
\mC^M}\left(\bdT_{m,n,p,q}^{[j]}\right)^{\alpha_{m,n,p,q,j}}\bdw_i\right \}
\label{eq.bfcolumnkgel.M}
\end{align}
where
\begin{align}
\alpha_{m,n,p,q,j}\in\begin{cases}\{(q-1)(l+1),(q-1)(l+1)+1,\dots,q(l+1)-1\}&\quad n>k\\
\{(q-1)(l+1),(q-1)(l+1)+1,\dots,q(l+1)-2\}&\quad n\leq k\\
\end{cases} \label{eq.alpha.mnpqj}
\end{align}
 \end{enumerate}

In summary, the beamforming design is as follows. For  message $K$,
we construct a beamforming column set as  \begin{align*} \td
\mV_k&\!=\bigcup_{1\leq q \leq M}\!\left\{\!
  \prod_{(m,n,p,q,j)\in \mC^M}\!
    \left(\bdT_{m,n,p,q}^{[j]}\right)^{\!\alpha_{m,n,p,q,j}}\!\!\bdw_i\bigg|
 1\leq i \leq \bar d_k\right\}
\end{align*}
where $\alpha_{m,n,p,q,j}$ satisfies \eqref{eq.alpha.mnpqj}.
The beamforming matrix $\td\bdV_k$ is chosen to be the matrix that
contains all the columns of $\td \mV_k$, which has $\bar d_kM l^{\Gamma_{k,q}^M}(l+1)^{\Gamma^M_q-\Gamma^M_{k,q}}$
columns.

\subsubsection{Alignment at the receivers}
Notice that the beamforming columns can be divided into $M$ parts based
on different values of $q$, which determines the range of the exponents that
associates with the $\bdT_{m,n,p,q}^{[j]}$ matrices. For any fixed value
of $q$, the proof of alignment at the receivers is the same as the single antenna case.

\subsubsection{Achievable Rates}

It is evident that $\td\bdV_k$ is a tall matrix of dimension
$\kappa M^2(l+1)^{\Gamma^M/M}\times \bar d_kM l^{\Gamma_{k,q}^M}(l+1)^{\Gamma^M_q-\Gamma^M_{k,q}}$. We can verify that it
has full column rank based on \cite[Lemma 1]{caja09}.  Therefore,  for each
antenna of transmitter $k$, the message has the following DoF
\begin{align*}
\lim_{l\rightarrow
\infty}\frac{|\td\bdV_k|}{\tau}=\lim_{l\rightarrow \infty}
\frac{\bar d_kM l^{\Gamma_{k,q}^M}(l+1)^{\Gamma^M_q-\Gamma^M_{k,q}}}{\kappa M^2(l+1)^{\Gamma^M/M}}
=\frac{\bar d_{k}}{\kappa M}=\frac{d_k}{M}.
\end{align*}
Notice that the channels $\td\bdh_{jk,p}, k\in\mM_j, 1\leq p \leq M$ are
linearly independent, therefore the messages from virtual single antenna
transmitters are orthogonal to each other. Hence, transmitter $k$  can send message with DoF $d_k$ as
it has   $M$ transmit antennas.

\subsubsection{Separation of the signal and interference spaces}

Finally, we need to ensure that the interference space and signal space are
linearly independent for all the receivers. This is similar to the proof in
single antenna case as well. For given value of $q$, the proof is the same. On
the other hand, the blocks associated with different $q$ are apparently linear
independent due to the non-overlapping range of exponents.

Hence, combining the interference alignment and the full-rank arguments, we conclude
that any point $\bdd$ satisfying \eqref{eq.main} is achievable for multiple
antenna case.

\section{Discussion} \label{sec.discuss} In this section, we outline some
alternative schemes that require a lower level of time-expansion for achieving
the same DoF region, and highlight some interesting consequences of the
general results developed in Section \ref{sec.general.multicast}.

\subsection{Group based alignment scheme} \label{sec.scheme.mod} The
achievability scheme presented in \secref{sec.general.multicast} requires all
interference messages at one receiver to be aligned with the largest one. This
may introduce more alignment constraints than needed. We give an example here
to illustrate this point.

\begin{figure}
\centerline{\includegraphics[width=0.80\figwidth]{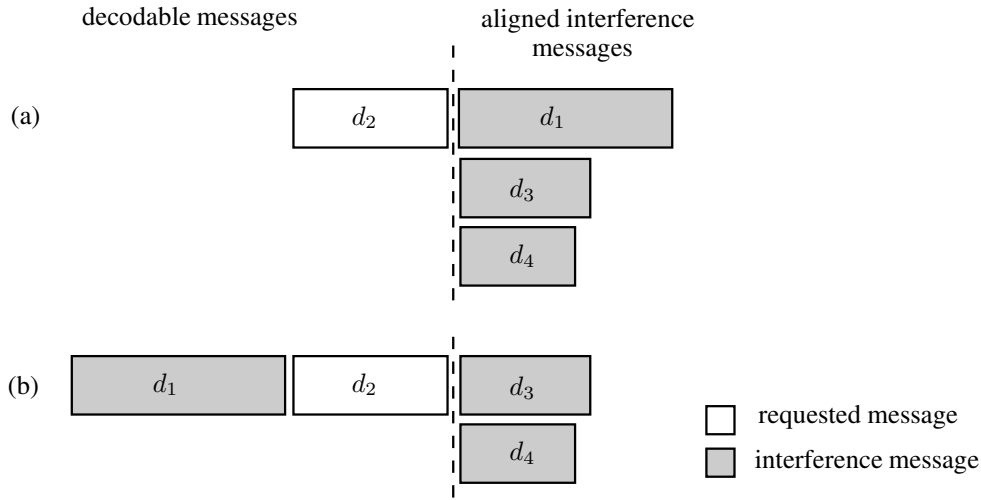}}
\caption{Example of alignment: (a) the original scheme, (b) the modified
scheme.} \label{fig.align}
\end{figure}

\begin{example}\label{exp.alternate} Consider a simple scenario where there are $4$ messages and
5 receivers. Without loss of generality, assuming \eqref{eq.relation.vi} is
true and $\mM_1=\{1,2\}$, $\mM_2=\{2\}$, $\mM_3=\{2,3\}$, $\mM_4=\{2,3\}$ and
$\mM_5=\{1,4\}$. The alignment constraints associated with the first two
receivers will be the following
\begin{align*}
\td\bdH_{14}\td\bdV_4 &\prec \td\bdH_{13} \td\bdV_3,\\
\td\bdH_{23}\td\bdV_3 &\prec \td\bdH_{21} \td\bdV_1, \\
\td\bdH_{24}\td\bdV_4 &\prec \td\bdH_{21} \td\bdV_1.
\end{align*}
However, in this particular case, upon inspection, one can realize that even
if receiver 2 also receives message 1, the DoF region will not change. This is
because the constraint at receiver 1 dictates that
\begin{align*}
d_1+d_2+\max(d_3,d_4)\leq M.
\end{align*}
However, this also implies the required constraint at receiver 2, which
is\begin{align*} d_2+\max(d_1,d_3,d_4)\leq M.
\end{align*}

Therefore, receiver 2 can use the same alignment relationship as receiver 1,
i.e., it can also decode message 1 without shrinking the DoF region. The
difference between the original alignment scheme and the modified scheme of
receiver 2 is illustrated in \figref{fig.align}. \hfill \QED
\end{example}

The alignment scheme of \secref{sec.general.multicast} can be modified
appropriately using the idea of \emph{partially ordered set}
(poset)\cite{bdhp02}.

A poset is a set $\mP$ and a binary relation $\leq$ such that for all $a, b, c
\in \mP$, we have
\begin{enumerate}
\item $a\leq a$ (reflexivity).
\item $a \leq b$ and $b \leq c$ implies $a \leq c$ (transitivity).
\item $a \leq b$ and $b \leq a$ implies $a = b$ (antisymmetry).
\end{enumerate}
An element $b$ in $\mP$ is the \emph{greatest element} if for every element $a
\in P$, we have $a \leq b$. An element $b \in P$ is a \emph{maximal element}
if there is no element $a \in \mP$ such that $a > b$. If a poset has a
greatest element, it must be the unique maximal element, but otherwise there
can be more than one maximal element.

For two message request sets $\mM_j$ and $\mM_{j'}$, we say $\mM_j \leq
\mM_{j'}$ if $\mM_j \subseteq \mM_{j'}$. With this partial ordering, the
collection of message request sets $\{\mM_j: 1\le j\le K \}$, with duplicate
elements (message sets) removed, forms a poset. Let $G$ denote the number of
maximal elements of this poset, and $\bar \mM_g$ denote the $g$th maximal
element, $1\le g\le G$. We divide the receivers into $G$ group according to
the following rule: For receiver $j$, if there exists a group index $g$ such
that $\mM_j=\bar\mM_g$, then receiver $j$ is assigned to group $g$. Otherwise,
$\mM_j$ is not a maximal element, we can assign receiver $j$ to any group $g$
such that $\mM_j\subset \bar\mM_g$. In the case where there are multiple
maximal elements of the poset that are ``larger'' than $\mM_j$, we can choose
the index of any of them as the group index of receiver $j$.

With our grouping scheme, there will be at least one receiver in each group
whose message request set is a superset of the message request set of any
other receiver in the same group. There may be multiple such receivers in each
group though. In either case, we term one such (or the one in case there is
only one) receiver as the \emph{prime} receiver. We choose all the receivers
within one group use the same alignment relationship as the prime receiver of
that group and the total number of alignment constraints is reduced. In such a
way, the receivers in one group can actually decode the same messages
requested by the prime receiver of that group, and they can simply discard the
messages that they are not interested in.

For instance in \expref{exp.alternate} given in this section, we can divide 5
receivers into three groups. Receivers 1 and 2 as group 1, receivers 3 and 4
as group 2, receiver 5 as group 3 and prime receivers are 1, 3 and 5. We
remark that there are multiple ways of group division as long as one receiver
can only belong to one group, e.g., receiver 1 as group 1, receivers 2, 3 and
4 as group 2, receiver 5 as group 3 and prime receivers are 1, 4 and 5.

In line with the above discussion, we have the following result.
\begin{cor} \label{cor.primerxs}
The DoF region of the interference network with general message requests as in
\secref{sec.model} is determined by the prime receivers. Adding non-prime
receivers to the system will not affect the DoF region.
\begin{IEEEproof} This can be shown as the inequalities \eqref{eq.main}
associated with the non-prime receivers are inactive, therefore the region is
dominated by the inequalities of prime receivers.
\end{IEEEproof}
\end{cor}
\subsection{DoF region of $K$ user $M$ antenna interference channel} As we
point out before, the $K$ user $M$ antenna interference channel is a special
case of the model we considered in this paper, hence, its DoF region can be
directly derived based on \thmref{thm.main}.
\begin{cor}
The DoF region of $K$ user $M$ antenna interference channel is
\begin{equation}\label{eq.dofic}
\mathcal{D} = \left\{
  \left( {{d_1},{d_2}, \cdots ,{d_K}} \right): \quad 0\leq{d_i}
+ {d_j} \le M, \forall 1 \le i,j \le K,i \ne j \right\}.
\end{equation}
\end{cor}
As a special case of our interference network with general message request,
the corollary requires no new proof. But we here give an alternative scheme
based on simple time sharing argument.
\begin{IEEEproof}
Without loss of generality, suppose $d_1^*\geq d_2^* \geq d_k^*,k = 3,\cdots,
K$, and $d_i^*+d_j^*\le d_1^*+d_2^*\le M$, $\forall i,j\in \{1,2,\dots K\}$.
We would like to show that $(d_1, d_2, \ldots, d_K)=(d_1^*, d_2^*, \ldots,
d_K^*)$ is achievable.

It is obvious that
\[
    (d_1, d_2, \ldots, d_K)=(M,0,\ldots, 0)
\]
can be achieved by single user transmission. It is also known from
\cite{caja08} that the point
\[
    (d_1, d_2, \ldots, d_K)=(M/2, M/2,\ldots, M/2)
\]
is achievable. Trivially, the point
\[
    (d_1, d_2, \ldots, d_K)=(0,0, \ldots, 0)
\]
is achievable.

By time sharing, with weights $(d_1-d_2)/M$, $2d_2/M$ and $1-d_1/M-d_2/M$
among the three points, in that order, it follows that the point
\[
(d_1, d_2, \ldots, d_K)=(d_1^*, d_2^*, d_2^*, \ldots, d_2^*)
\]
is achievable. This is already at least as large as the DoF we would like
to have.
\end{IEEEproof}

\textbf{Remark}: After the submission of our manuscript the following results have appeared that are related to our work. The DoF region for a single-antenna interference channel without
time-expansion has been shown to be the convex hull of
$\{\bde_1, \ldots, \bde_K, \frac 12 \bdones \}$ for almost all (in Lebesgue
sense) channels \cite{wusv11i}. Interestingly, this agrees with DoF region of
the $K$-user single antenna interference channel. For, it can be seen from the
proof of Corollary~\ref{cor.primerxs} that the DoF region given in
\eqref{eq.dofic} can be alternatively formulated as the convex hull of the
vectors $\{\bdzeros, M \bde_1, \ldots, M\bde_K, \frac M2 \bdones\}$. Setting
$M=1$ will yield the desired equivalence of the two DoF regions. This
equivalence, is non-trivial, however, because it shows that allowing for
time-expansion, and time-diversity (channel variation), the DoF region of the
interference channel is not increased --- the DoF is an inherent spatial (as
opposed to temporal) characteristic of the interference channel.

\subsection{Length of time expansion}

For the $K$ user $M$ antenna interference channel, the total length of time
expansion needed in \cite{caja08} is smaller than our scheme in order to
achieve $KM/2$ total DoF. This is due to the fact that when $J=K$ and
$\mM_j=\{j \},\forall j$, it is possible to choose $\td \bdV_2$ carefully such
that the cardinality of $\td \bdV_2$ is the same as $\td \bdV_1$ and there is
one-to-one mapping between these two. For other asymmetric DoF points, it is
in general not possible to choose two messages having the same cardinality of
beamforming column sets. The total time expansion needed could be reduced if
we use the group based alignment scheme in \secref{sec.scheme.mod} and/or
design the achievable scheme for a specific network with certain DoF.

\subsection{The total DoF of an interference network with general message
demands} As a byproduct of our previous analysis, we can also find the total
degrees of freedom for an interference network with general message demands.
\begin{cor} The total DoF of an interference network with general message
demands can be obtained by a linear program shown as follows
\begin{align}
& \max\sum_{k=1}^K d_k  \nonumber\\
& \text{s.t.}\quad \sum_{k\in \mM_j}d_k+\max_{i\in\mM_j^c}(d_i)\leq M,\quad
  \forall 1\leq j\leq J, \bdd\in \mathbb{R}^K_+ .\label{eq.lp.constraint}
\end{align} \hfill \QED
\end{cor}

\begin{cor}
If all prime receivers demand $\beta$, $1\leq \beta \leq K-1$, messages, and
each of the $K$ messages is requested by the same number of prime receivers.
Then the total DoF is
\begin{align}
d_\tot=\frac{MK}{\beta+1}, \label{eq.dofsum.alpha}
\end{align}
 and is achieved by \begin{align}
\bdd=\left(\frac{M}{\beta+1},\frac{M}{\beta+1},\dots,\frac{M}{\beta+1}\right).
\label{eq.dofvec.alpha}
\end{align}
\end{cor}
\begin{IEEEproof}
Based on \corref{cor.primerxs}, we only need to consider $G$ inequalities
(where $G$ is the number of groups) that are associated with the prime
receivers. We show that \eqref{eq.dofvec.alpha} achieves the maximum total DoF
when all $K$ messages are requested by the same number of prime receivers.
Notice that in this case we can expand the inequality of
\eqref{eq.lp.constraint} into $K-\beta$ inequalities by removing the $\max()$
operation. Hence, we will have $G(K-\beta)$ inequalities in total. Since each
message is requested by $G\beta/K$ prime receivers, for each $d_k$ it appears
$\frac{G\beta }{K} (K-\beta)$ times among the inequalities for prime receivers
which request $d_k$, and it appears $G-\frac{G\beta}{K}$ times otherwise.
Summing all the $G (K-\beta))$ inequalities we have
\begin{align*}
\left(\frac{G \beta }{K} (K-\beta)+G-\frac{G\beta}{K}\right)\sum_kd_k\leq
MG(K-\beta).
\end{align*}
Hence
\begin{align*}
\sum_kd_k\leq \frac{MK}{\beta+1},
\end{align*}
and the corollary is proven.
\end{IEEEproof}

\begin{remark}
If messages are not requested by the same number of prime receivers it is
possible to achieve a higher sum DoF than \eqref{eq.dofsum.alpha}. We only
need to show an example here. Assuming that there are $4$ transmitters and $3$
prime receivers, the message requests are $\{1,2\},\{1,3\},\{1,4\}$. If all
the transmitters send $M/3$ DoF, we could achieve \eqref{eq.dofsum.alpha}.
However, choosing $\bdd=(0,\frac{M}{2},\frac{M}{2},\frac{M}{2})$ will lead to
sum DoF $3M/2$ which is higher.
\end{remark}

\section{Conclusions and Future Work} \label{sec.con} We derived the DoF
region of an interference network with general message demands. The region is
a convex polytope, which is the intersection of a number of cylindrical sets
whose projections into lower dimensions are simple geometric shapes each
enclosed by a simplex and the coordinate planes. In certain special cases, it
is possible to find the vertices of the DoF region polytope explicitly. One
such case is the $K$-user $M$-antenna interference channel with multiple
unicasts, whose DoF region is a convex hull of simple points of the all zero
vector, the scaled natural basis vectors, and a scaled all-one vector, which
interestingly coincides with the DoF region recently obtained for
Lebesgue-a.e.\ constant coefficient channels with no time diversity.

Our achievability scheme for deriving the DoF region operates by generating
beamforming columns with multiple base vectors over time expanded channel, and
aligning the interference at each receiver to its largest interferer. We also
showed that the DoF region is determined by a subset of receivers (called
prime receivers), that can be identified by examining the message demands of
the receivers. We provided an alternate interference alignment scheme in this
scenario, where the certain receivers share the same alignment relationship,
which helps to reduce the required duration of for time-expansion.

It would be interesting to consider general message demands in other
interference networks. For instance, if each transmitter has multiple
messages, the receiver demands may result in alignment constraints that cannot
be satisfied in the same manner as described in this paper. On the other hand,
the usage of multiple base vectors may be useful in proving achievability for
other problems where interference alignment is applicable.

\linespread{1.5}


\begin{thebibliography}{10}
\providecommand{\url}[1]{#1}
\def\UrlFont{\rmfamily}
\providecommand{\newblock}{\relax} \providecommand{\bibinfo}[2]{#2}
\providecommand\BIBentrySTDinterwordspacing{\spaceskip=0pt\relax}
\providecommand\BIBentryALTinterwordstretchfactor{4}
\providecommand\BIBentryALTinterwordspacing{\spaceskip=\fontdimen2\font
plus \BIBentryALTinterwordstretchfactor\fontdimen3\font minus
  \fontdimen4\font\relax}
\providecommand\BIBforeignlanguage[2]{{%
\expandafter\ifx\csname l@#1\endcsname\relax
\typeout{** WARNING: IEEEtran.bst: No hyphenation pattern has been}%
\typeout{** loaded for the language `#1'. Using the pattern for}%
\typeout{** the default language instead.}%
\else \language=\csname l@#1\endcsname \fi #2}}
\bibitem{mamk06c}
M.~Maddah-Ali, A.~Motahari, and A.~Khandani, ``Signaling over {MIMO}
  {Multi-Base} {Systems:} {Combination} of {Multi-Access} and {Broadcast}
  {Schemes},'' in \emph{Proc. IEEE Intl. Symp. on Info. Theory}, 2006, pp.
  2104--2108.

\bibitem{jash08}
S.~Jafar and S.~Shamai, ``Degrees of {freedom} {region} of the
{MIMO} {$X$}
  {channel},'' \emph{{IEEE} Trans. Inform. Theory}, vol.~54, no.~1, pp.
  151--170, Jan. 2008.

\bibitem{caja08}
V.~Cadambe and S.~Jafar, ``Interference {alignment} and {degrees} of
{freedom}
  of the {$K$} user {interference} {channel},'' \emph{{IEEE} Trans. Inform.
  Theory}, vol.~54, no.~8, pp. 3425--3441, Aug. 2008.

\bibitem{caja09}
V.~Cadambe and S.~Jafar, ``Interference {Alignment} and the
{Degrees} of
  {Freedom} of {Wireless} {$X$} {Networks},'' \emph{{IEEE} Trans. Inform.
  Theory}, vol.~55, no.~9, pp. 3893--3908, Sept. 2009.

\bibitem{suts08c}
C.~Suh and D.~Tse, ``Interference {Alignment} for {Cellular}
{Networks},'' in
  \emph{Proc. of Allerton Conf. Commun., Control, and Computing}, 2008, pp.
  1037--1044.

\bibitem{wesk07c}
H.~Weingarten, S.~Shamai, and G.~Kramer, ``{On the compound MIMO
broadcast
  channel},'' in \emph{Proceedings of Annual Information Theory and
  Applications Workshop UCSD}, 2007.

\bibitem{gojw09}
T.~Gou, S.~Jafar, and C.~Wang, ``On the {Degrees} of {Freedom} of
{Finite}
  {State} {Compound} {Wireless} {Networks},'' \emph{{IEEE} Trans. Inform.
  Theory}, vol.~57, no.~6, pp. 3286--3308, June 2011.

\bibitem{mali09a}
\BIBentryALTinterwordspacing M.~A. Maddah-Ali, ``{On the Degrees of
Freedom of the Compound MIMO Broadcast
  Channels with Finite States},'' 2009. [Online]. Available:
  \url{http://arxiv.org/abs/0909.5006}
\BIBentrySTDinterwordspacing

\bibitem{brpt10}
G.~Bresler, A.~Parekh, and D.~N.~C. Tse, ``The approximate capacity
of the
  many-to-one and one-to-many gaussian interference channels,'' \emph{{IEEE}
  Trans. Inform. Theory}, vol.~56, no.~9, pp. 4566--4592, Sept. 2010.

\bibitem{sjvj08c}
S.~Sridharan, A.~Jafarian, S.~Vishwanath, S.~A. Jafar, and
S.~Shamai, ``A
  layered lattice coding scheme for a class of three user gaussian interference
  channels,'' in \emph{46th Annual Allerton Conference Control, and Computing,
  on Communication}, 2008, pp. 531--538.

\bibitem{etor09}
\BIBentryALTinterwordspacing R.~H. Etkin and E.~Ordentlich, ``On the
degrees-of-freedom of the {$K$}-user
  gaussian interference channel,'' 2009. [Online]. Available:
  \url{http://arxiv.org/abs/0901.1695}
\BIBentrySTDinterwordspacing

\bibitem{mgmk09}
\BIBentryALTinterwordspacing A.~S. Motahari, S.~O. Gharan, M.~A.
Maddah-Ali, and A.~K. Khandani, ``Real
  interference alignment: Exploiting the potential of single antenna systems,''
  2009. [Online]. Available: \url{http://arxiv.org/abs/0908.2282}
\BIBentrySTDinterwordspacing

\bibitem{njgv09c}
B.~Nazer, S.~Jafar, M.~Gastpar, and S.~Vishwanath, ``Ergodic
interference
  alignment,'' in \emph{Proc. IEEE Intl. Symp. on Info. Theory}, 2009, pp.
  1769--1773.

\bibitem{ngjv09c}
B.~Nazer, M.~Gastpar, S.~Jafar, and S.~Vishwanath, ``Interference
alignment at
  finite {SNR:} {General} message sets,'' in \emph{Proc. of Allerton Conf.
  Commun., Control, and Computing}, 2009, pp. 843--848.

\bibitem{jafa07}
S.~Jafar and M.~Fakhereddin, ``Degrees of {freedom} for the {MIMO}
  {interference} {channel},'' \emph{{IEEE} Trans. Inform. Theory}, vol.~53,
  no.~7, pp. 2637--2642, July 2007.

\bibitem{goja08a}
T.~Gou and S.~Jafar, ``Degrees of {Freedom} of the {$K$} {User} {$M
\times N$}
  {MIMO} {Interference} {Channel},'' \emph{{IEEE} Trans. Inform. Theory},
  vol.~56, no.~12, pp. 6040--6057, Dec. 2010.

\bibitem{bdhp02}
B.~A. Davey and H.~A. Priestley, \emph{Introduction to lattices and
  order}.\hskip 1em plus 0.5em minus 0.4em\relax Cambridge University Press,
  2002.

\bibitem{wusv11i}
Y.~Wu, S.~Shamai~(Shitz), and S.~Verd\'{u}, ``Degrees of freedom of
the
  interference channel: a general formula,'' in \emph{Proc. IEEE Intl. Symp. on
  Info. Theory}, 2011, pp. 1344--1348.
\end{thebibliography}
\end{document}